\documentclass[superscriptaddress,twocolumn,prl]{revtex4}

\usepackage{graphicx}
\usepackage{dcolumn}
\usepackage{multirow}
\usepackage{bm}
\usepackage{times}

\usepackage{color}
\usepackage{amstext}
\usepackage{amsmath}
\usepackage{amsfonts}

\newcommand{\etal}{\textit{et al.}}
\newcommand{\BKFA}{$\rm Ba_{0.68}K_{0.32}Fe_2As_2$}
\newcommand{\BKFAX}{$\rm Ba_{1-x}K_{x}Fe_2As_2$}
\newcommand{\BFCA}{$\rm Ba(Fe_{0.925}Co_{0.075})_2As_2$}
\newcommand{\BFCAX}{$\rm Ba(Fe_{1-x}Co_x)_2As_2$}
\newcommand{\BFMA}{$\rm Ba(Fe_{0.88}Mn_{0.12})_2As_2$}
\newcommand{\BFA}{$\rm BaFe_2As_2$}
\newcommand{\KFA}{$\rm KFe_2As_2$}
\newcommand{\mJ}{mJ/molK$^2$}
\usepackage{amssymb}

\begin{document}

\title{Specific heat of $\bf Ba_{0.68}K_{0.32}Fe_2As_2$: evidence for multiband strong-coupling
superconductivity}

\author{P. Popovich}
\affiliation{Max-Planck-Institut f\"{u}r Festk\"{o}rperforschung,
Heisenbergstrasse 1, D-70569 Stuttgart, Germany}
\author{A.V. Boris}
\affiliation{Max-Planck-Institut f\"{u}r Festk\"{o}rperforschung, Heisenbergstrasse 1, D-70569 Stuttgart, Germany}
\affiliation{Department of Physics, Loughborough University, Loughborough,
LE11 3TU, United Kingdom}
\author{O.V. Dolgov}
\affiliation{Max-Planck-Institut f\"{u}r Festk\"{o}rperforschung,
Heisenbergstrasse 1, D-70569 Stuttgart, Germany}
\author{A.A. Golubov}
\affiliation{Faculty of Science and Technology, University of Twente, 7500 AE Enschede, The Netherlands}
\author{D.L. Sun}
\affiliation{Max-Planck-Institut f\"{u}r Festk\"{o}rperforschung,
Heisenbergstrasse 1, D-70569 Stuttgart, Germany}
\author{C.T. Lin}
\affiliation{Max-Planck-Institut f\"{u}r Festk\"{o}rperforschung,
Heisenbergstrasse 1, D-70569 Stuttgart, Germany}
\author{R.K. Kremer}
\affiliation{Max-Planck-Institut f\"{u}r Festk\"{o}rperforschung,
Heisenbergstrasse 1, D-70569 Stuttgart, Germany}
\author{B. Keimer}
\affiliation{Max-Planck-Institut f\"{u}r Festk\"{o}rperforschung, Heisenbergstrasse 1, D-70569 Stuttgart, Germany}
\date{\today}
\begin{abstract}
The specific heat of high-purity $\rm Ba_{0.68}K_{0.32}Fe_2As_2$ single crystals with the highest reported superconducting $T_c$ = 38.5 K was studied. The electronic specific heat, $C_p$, below $T_c$ shows two gap features, with $\Delta_1 \approx 11$ meV and $\Delta_2 \approx 3.5$ meV obtained from an $\alpha$-model analysis. The reduced gap value, $2\Delta^{\rm max} / k_B T_c \approx 6.6$, the magnitude of the specific heat jump, $\Delta C_p(T_c)/T_c$, and its slope below $T_c$ exhibit strong-coupling character. We also show that an Eliashberg model with two hole and two electron bands gives the correct values of $T_c$, the superconducting gaps, and the temperature dependence of the free-energy difference.
\end{abstract}


\maketitle

The newly discovered iron pnictide superconductors present an unusual case of multiband superconductivity in the vicinity of a magnetic instability. In view of the weak electron-phonon coupling in this class of superconductors \cite{Lilia}, magnetic excitations are the most promising candidates for the pairing boson. The analysis of experimental data and the theoretical modeling of these systems are, however, considerably complicated by their complex electronic structure that involves at least four energy bands crossing the Fermi surface. Despite intensive research, there is hence no consensus on the gap symmetry and on the nature of the pairing interaction.

The thermodynamic properties are a key source of surface-insensitive
information on the electronic interactions and on the structure of the superconducting gap function. For a superconducting order parameter with alternating sign, which appears to be present in at least some of the pnictides, the pairing interactions are strongly influenced by impurity scattering, so that sample quality and phase purity issues are crucial for calorimetric experiments. Presumably because of such issues, the current specific heat evidence of  doped \BFA, a pnictide family for which sizable single crystals are available, is still ambiguous. Recently two gaps were reported in electron-doped \BFCA\ , supporting the results of nuclear magnetic resonance (NMR) and muon spin relaxation ($\mu$SR) studies \cite{Hardy}. In hole-doped \BKFAX, on the other hand, prior specific heat experiments have shown only one gap with values of 6 meV in a single crystal with $T_c$ = 36.5 K \cite{Mu} and varying from 5.9 meV to 6.7 meV for 0.3 $\leq x \leq$ 0.6 (36.0 K $\leq T_c \leq$ 37.3 K) in polycrystalline specimens \cite{Kant}, in contrast to the observation of two nearly isotropic gaps in angle-resolved photoemission (ARPES) experiments \cite{Evt}. Alas, all of the specific heat data thus far reported suffer from a residual low-temperature non-superconducting electronic contribution and show Schottky anomalies. In electron-doped \BFA, this complication is likely due to structural defects caused by Co substitution directly in the superconducting FeAs planes. In hole-doped \BFA, the substitution of Ba by K does not disturb the structure of the FeAs planes, but gives rise to electronic phase separation in the underdoped regime \cite{Park,Inosov}.

In this Letter we report a comprehensive study of the specific heat, $C_p$, of an optimally hole-doped, high-purity \BKFA\ single crystal with $T_c$ = 38.5 K (onset), the highest transition temperature reported to-date for this family of superconductors. The superconducting specific heat indicates two energy gaps with magnitudes $2\Delta_1 = 6.6\ k_B T_c$ and $2\Delta_2 = 2.2\ k_B T_c$, in agreement with the ARPES, $\mu$SR, and NMR results \cite{Evt,Matano}.
We also calculated the specific heat from the Eliashberg spectral function in a four-band spin-fluctuation model. From the solution of the Eliashberg equations we obtained the value of $T_c$, the superconducting gaps, and the temperature dependence of the free-energy difference in good agreement with the experimental data. The averaged electron-boson coupling constant, $\lambda^{av} \approx 1.9$, resulting from this analysis and consistent with the magnitude of the specific heat jump and its first derivative below $T_c$, is distinctly larger than the estimated electron-phonon coupling \cite{Lilia}. The strong coupling suggests the interaction with low-energy ($\lesssim$ 50 meV) interband spin-fluctuations as the pairing mechanism.

\begin{figure}[ht]
\includegraphics[width=7.7cm]{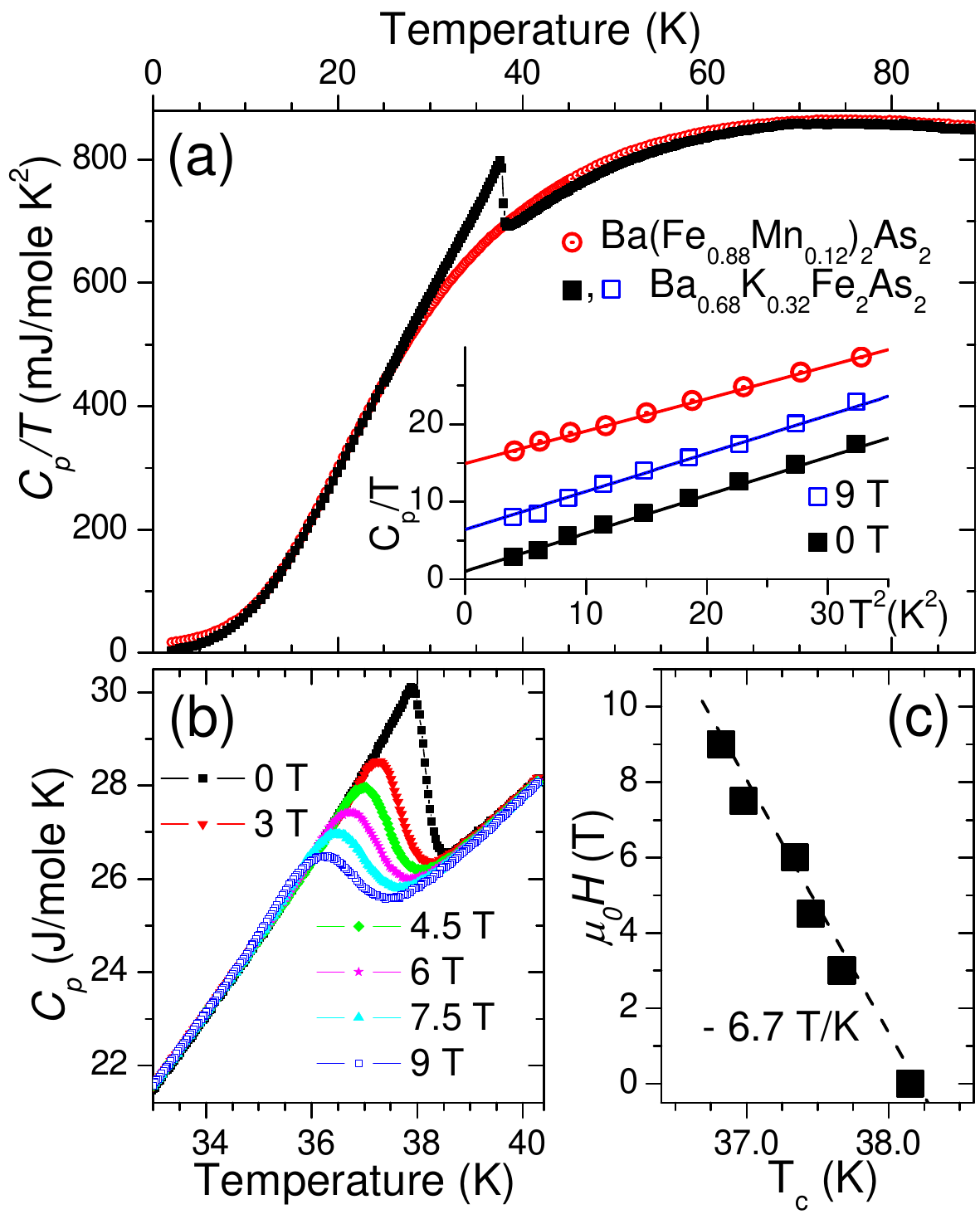}
\caption{(a) Temperature dependence of the specific heat $C_p/T$ of \BKFA\ (squares) and \BFMA\ (circles). The inset shows a plot
of $C_p/T$ vs. $T^2$ at low $T$ for both samples. The lines represent the best fit to $C_p(T)/T = \gamma (0) + \beta T^2$. (b,c) Temperature dependence of $C_p$ of \BKFA\ near $T_c$ measured at different external magnetic fields applied perpendicular to the $ab$-plane.}
\label{fig1}
\end{figure}

The superconducting \BKFA\ and reference \BFMA\ single crystals were grown from self-flux in zirconia crucibles sealed in quartz ampoules under argon atmosphere, as described earlier \cite{Sun}.
Their chemical compositions were determined by energy-dispersive X-ray spectrometry. The specific heat was measured in the temperature range between 2 and 200 K with a Physical Properties Measurement System (Quantum Design) using the thermal relaxation technique. The samples were mounted on a standard puck using Apiezon N grease. The measurements showed no difference between several cleaved pieces from the same batches. Bulk superconductivity in \BKFA\ was confirmed by magnetic susceptibility, $dc$ and infrared conductivity measurements, whereas \BFMA\ did not show any signatures of superconducting or magnetic phase transitions. In the following, we present specific-heat data for a superconducting \BKFA\ crystal of weight 13.6 mg and for a \BFMA\ reference crystal of weight 9.9 mg.

Figure 1a displays unprocessed specific heat data on the \BKFA\ and \BFMA\ single crystals. $C_p(T)/T$ of \BKFA\ shows no signs of low-temperature upturns or Schottky anomalies at any magnetic field. In the low-temperature limit (inset in Fig. 1a), the data can be fitted by $C(T)/T = \gamma (0) + \beta T^2$, where $\gamma (0)T$ represents the residual zero-temperature electronic specific heat and the second term the lattice specific heat. The fitting parameters for zero field are $\gamma(0)=1.2(2)$ mJ/mol K$^2$ and $\beta$=0.496(1) mJ/mol K$^4$. For magnetic field H = 9 T we used the same $\beta$ and obtained $\gamma(0) = 6.4(2)$ mJ/mol K$^2$. These residual specific heat values are the lowest reported for FeAs-based superconductors so far, indicating the superior quality and high purity of our samples. The ratio of the residual electronic specific heat to its normal-state counterpart ($\gamma^N_{K}$, see below) yields an estimate of the non-superconducting phase fraction $\gamma(0)/\gamma_{K}^N\approx 1.2/50 = 2.4$ \%.
\begin{figure}[ht]
\includegraphics[width=7.7cm]{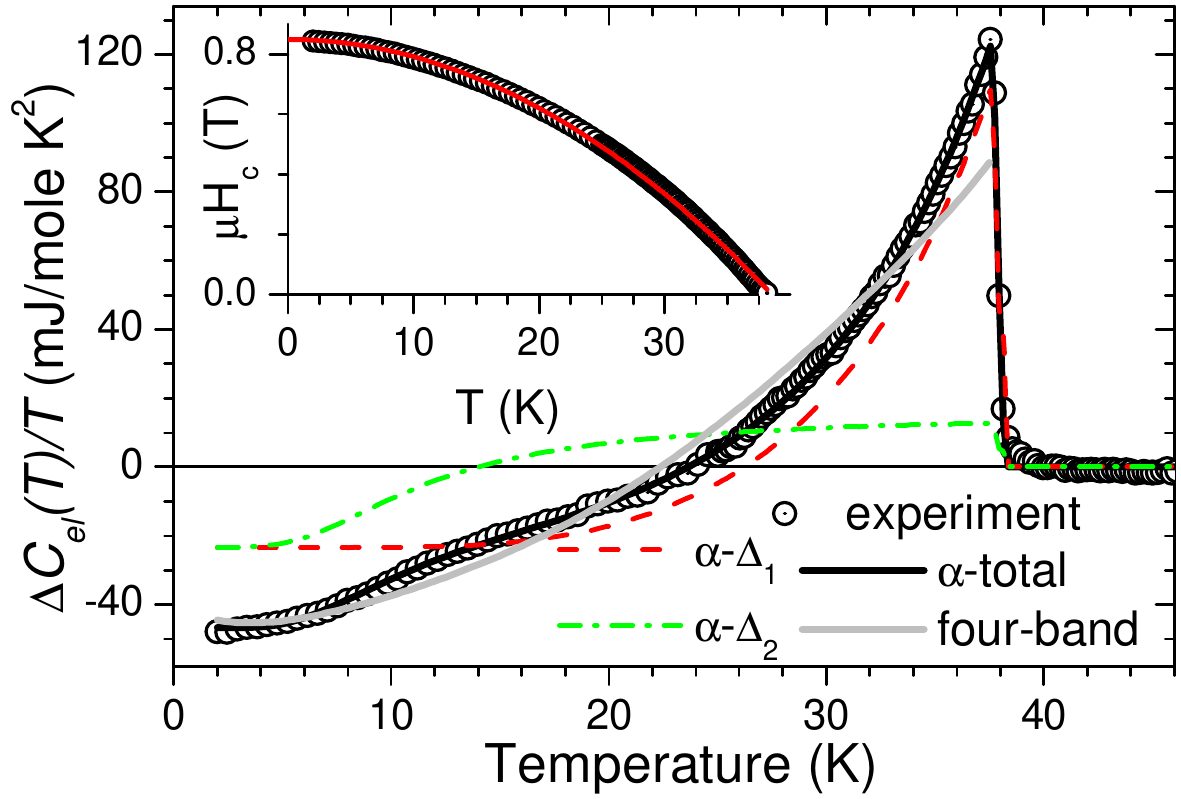}
\caption{ Superconductivity-induced specific-heat difference of \BKFA. The black solid curve is the result of a fit according
to the two-gap $\alpha$-model. The dashed and dash-dotted curves
represent the partial contributions of the two bands. The light gray line is the electronic specific heat calculated from the four-band Eliashberg
model. Inset: Temperature dependence of the thermodynamic critical
field  (see text).}
\label{fig2}
\end{figure}

Figure 1b highlights the superconductivity-induced jump anomaly of $C_p$, which exhibits the largest magnitude (125 mJ/mol K$^2$) and smallest width ($\leq 0.4$ K) reported to-date \cite{Hardy,Mu,Welp,Kant,Storey}. An external magnetic field applied perpendicular to the $ab$-plane gradually suppresses the jump and reduces the transition temperature by 1.4 K for $H=9$ T. The transition temperature $T_c$ determined as the maximum of the derivative of $C_p(T)$ decreases linearly with magnetic field with a slope of $\delta H_{c2}(T)/\delta T|_{T_{c}}$= - 6.7 T/K (Fig. 2c), that is consistent with the value reported in Ref. \onlinecite{Welp}. This is a factor of two larger than the value obtained from resistivity measurements on samples from the same batch \cite{Sun2}. This difference may be due to flux-flow effects.

In order to reliably extract the electronic contribution, $C_{el}(T)$, from the total measured specific heat, the contribution of lattice excitations, $C_{latt}(T)$, has to be accurately determined. 
The lattice specific heat of the parent compound \BFA\ can not be accurately obtained because of the magnetic and structural phase transitions at $\sim 140$ K. We found that substitution of 12\% Fe by Mn suppresses the spin-density-wave state and does not induce superconductivity. The specific heat of \BFMA\ (red circles in Fig. 1a) was found to be independent of magnetic field. The electronic and lattice specific heat terms extracted from a fit to low-temperature data on this compound (inset in Fig. 1a) are $\gamma(0)=\gamma^N_{Mn}$ = 14.9(2) mJ/mol K$^{2}$ and $\beta_{Mn} $ = 0.420(4) mJ/mol K$^{4}$, respectively. The lattice specific heat of \BFMA, $C_{latt}^{Mn}(T)$, was obtained by subtracting $\gamma^N_{Mn} T$ from its total specific heat, $C_{tot}^{Mn}(T)$. The estimated Debye temperature is 306 K, somewhat higher than the corresponding value for superconducting \BKFA\ (277 K). Since the difference of the lattice parameters of both compounds does not exceed 1.5\%, we assume that the phonon contributions to their specific heat and to the entropy obey a law of corresponding states. The normal-state  specific heat of \BKFA\ can then be obtained from the commonly used corresponding states approximation \cite{Stout}:
\begin{equation}
C_{tot}^{K}(T)=C_{latt}^{K}(T)+\gamma^N_{K}\cdot T=A\cdot C_{latt}^{Mn}(B\cdot T)+\gamma^N_K\cdot T
\end{equation}%
where $\gamma^N_{K}$ is the Sommerfeld constant in the normal state of \BKFA, and $A$ and $B$ are close to unity.
From a least-squares fit of our data to Eq. (1) within the temperature range 40 - 150 K (more than 300 data points) under the constraint of entropy conservation, we obtained $\gamma^N_{K}$= 50 mJ/mol K$^{2}$, with $A = 0.95$ and $B = 1.03$.
Figure 2 shows the difference, $\Delta C_{el}(T)/T$, between the measured specific heat $C_p(T)/T$ of \BKFA\ (Fig. 1a) and its normal-state counterpart $C_{tot}^{K}(T)/T$ estimated from Eq. (1). The same $\Delta C_{el}(T)/T$ (within the symbol size in Fig. 2) has also been obtained by representing the lattice contribution on a basis of Einstein modes \cite{EPAPS}. The value of the Sommerfeld constant $\gamma^N_{K}$= 50 mJ/mol K$^{2}$ is in between those reported for the pristine end compounds \BFA\ (6.1 mJ/molK$^{2}$ \cite{Sefat}) and \KFA\ (69.1 mJ/molK$^{2}$ \cite{Fukazawa}). This value exceeds the one calculated from the band structure ($\gamma_{DFT} \approx$ 10.1 mJ/molK$^{2}$ for $\rm Ba_{0.6}K_{0.4}Fe_2As_2$ \cite{Sasha}) by almost a factor of five, while it is consistently inferred from all specific heat experiments reported to date \cite{Mu,Kant,Storey}.

The thermodynamic critical field, $H_{c}(T)$, can be obtained from the free-energy difference between the normal and superconducting states:
\begin{equation}
\Delta F(T)=\mu _{0}V_{m}H_{c}^{2}(T)/2=\Delta U(T)-T\Delta
S(T)  \label{CriticalField}
\end{equation}
where $V_{m} = 6.05\times 10^{-5}$ m$^{3}$/mol is the molar volume determined from our X-ray diffraction measurements on \BKFA. The temperature dependence of the thermodynamic critical field obtained by numerical integration of our data is displayed in the inset of Fig. 2. By fitting the thermodynamic critical field expressed as $H_{c}(T)=H_{c0}\left( 1-\left( T/T_{c}\right) ^{2}\right) $ to our data between 2 and 38.5 K we obtained $\mu _{0}H_{c0}$ = 0.85 T, a value comparable to those of the cuprate superconductors \cite{Liang}. 

In contrast to $\gamma^N_{K}$, the superconductivity-induced electronic specific heat is very sensitive to the sample quality and phase purity. We now show that if the impurity scattering is minimized,
 $\Delta C_{el}(T)$ reveals the intrinsic multigap strong-coupling nature of the superconducting state. Two quantities are commonly used to assess the pairing strength, namely the reduced jump anomaly, $\Delta C_{el}(T_c)/(\gamma^N_{K}T_c)$, and the normalized slope of $\Delta C_{el}(T)$ right below $T_c$,
$g (T_c) = - \left. \partial [\Delta C_{el}(T)/(\gamma^N_{K}T_c)]/\partial T \right.| _{T_{c}}$. Their magnitudes in \BKFA, 2.5 and 11.1, respectively, are significantly larger than the corresponding BCS values of 1.43 and 3.77. In the framework of the Eliashberg formalism, the coupling strength can be expressed in terms of the ratio $\hbar \omega _{ln}/ k_B T_{c}$, where $\omega _{ln}$ is the logarithmically averaged frequency of bosons mediating the Cooper pairing \cite{Carbotte}. In Figure 3, we place the reduced superconducting jump and $g$-value of \BKFA\ on a universal plot \cite{Carbotte} as a function of this dimensionless coupling parameter. Both values correspond to $\hbar \omega _{ln}/ k_B T_{c} = 7.5$, close to the prototypical strong-coupling superconductor Pb$_{0.8}$Bi$_{0.2}$. This comparison indicates strong-coupling superconductivity in \BKFA.

\begin{figure}[ht]
\includegraphics[width=7.7cm]{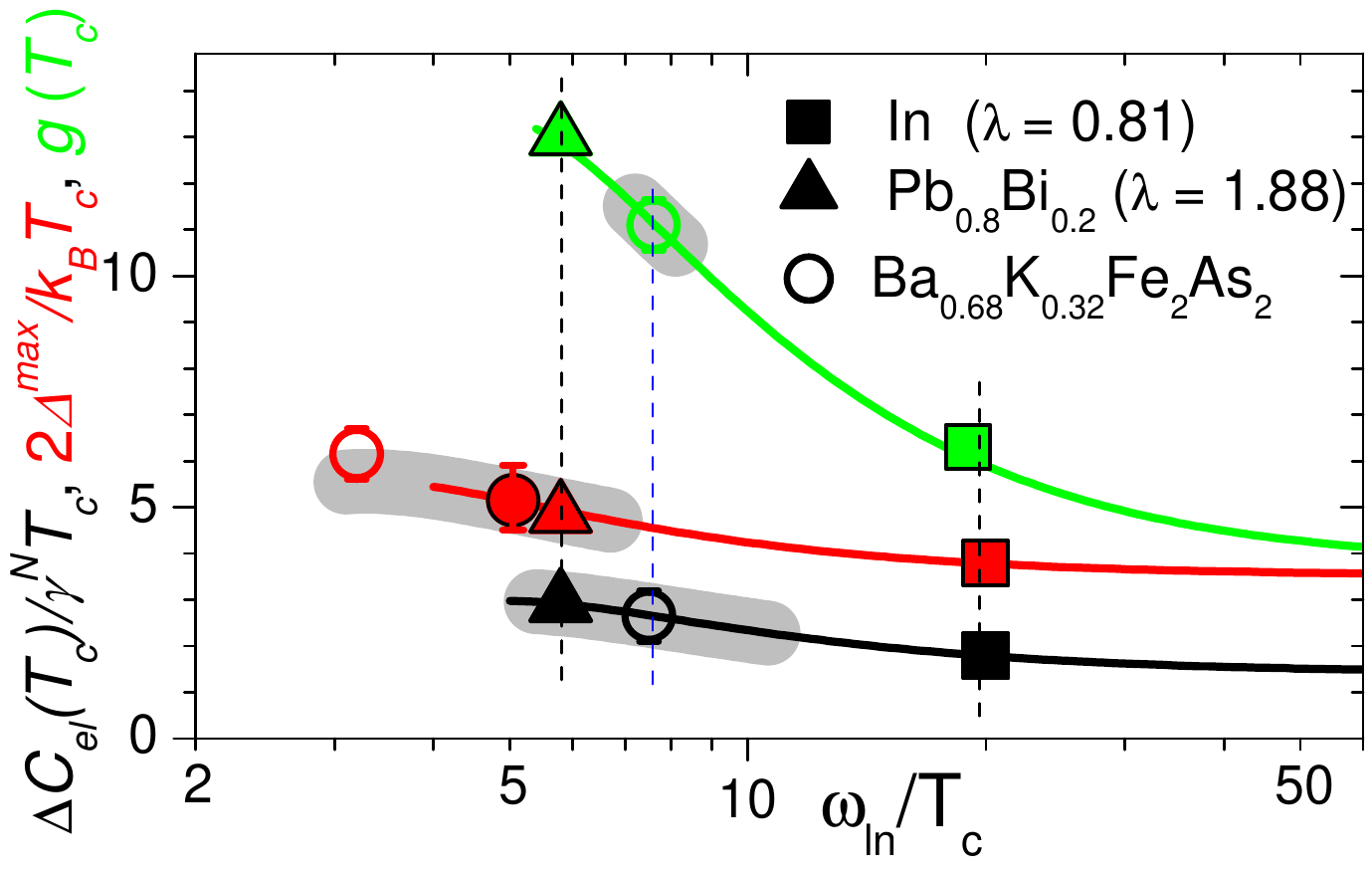}
\caption{Comparison of the dimensionless superconducting jump, energy gap at $T=0$, and slope of the specific heat near $T_c$ calculated by the semiphenomenological expressions in Ref. \onlinecite{Carbotte} (lines) with our results for \BKFA\ (open circles), and for Pb$_{0.8}$Bi$_{0.2}$ (triangles) and In (squares), after Ref. \onlinecite{Carbotte}. The open and  solid circles represent the maximum gap values obtained by the two-band $\alpha$ and four-band Eliashberg model, respectively.   Shaded areas represent the uncertainty in $\omega_{ln}/T_c$ according to the error bars in our data points. }
\label{fig3}
\end{figure}

The same issue can be addressed by considering the magnitude of the superconducting energy gap. Clearly, the prominent knee in the $\Delta C_{el}(T)/T$ data around $T=15$ K (Fig. 2) cannot be described in terms of models with a single gap. We have hence fitted these data to the phenomenological multi-band $\alpha$-model \cite{Padamsee}, which assumes a BCS temperature dependence of the gaps and has been widely used to analyze heat capacity data on $\rm MgB_2$ \cite{Dolgov}. The superconducting gap magnitudes at $T = 0$ are introduced as adjustable parameters $\alpha_j$, defined according to $\Delta_j(T)\,=\,(\alpha_j\,/\,\alpha_{\mathrm{BCS}})\,\Delta_{\mathrm{BCS}}$, where $\alpha_{\mathrm{BCS}} =\Delta_{\mathrm{BCS}}(0) / k_B T_c = 1.764$ is the weak-coupling value of the gap ratio. Another set of adjustable parameters is the fractions of the total electronic density of states, $\gamma_j/\gamma^N_K$, that each band contributes to the superconducting condensate. Figure 2 shows the result of our multigap fit (black solid curve) with only two gaps survived $\alpha_1$ = 3.3, $\alpha_2$ = 1.1, and $\gamma_1 \sim \gamma_2 \sim 0.5\cdot \gamma^N_K$,
which  reproduces well $\Delta C_{el}(T)/T$ below $T_c$. One of the gaps, $\Delta_2 (0)$ = 3.5 meV, is somewhat smaller than $\Delta_{\mathrm{BCS}}(0)$, whereas the other one, $\Delta_1 (0)$ = 11 meV, is much larger. Our analysis thus yields thermodynamic evidence of two different superconducting gaps (or groups of gaps) with quite different absolute values in \BKFA, in agreement with ARPES and
$\mu$SR results \cite{Evt}. Further, marking the corresponding point, $2\Delta^{\rm max}/k_B T_c$ = 6.6, on the universal plot of Fig. 3 strongly supports our conclusion about the strong-coupling nature of superconductivity in this compound.

\begin{figure}[ht]
\includegraphics[width=7.78cm]{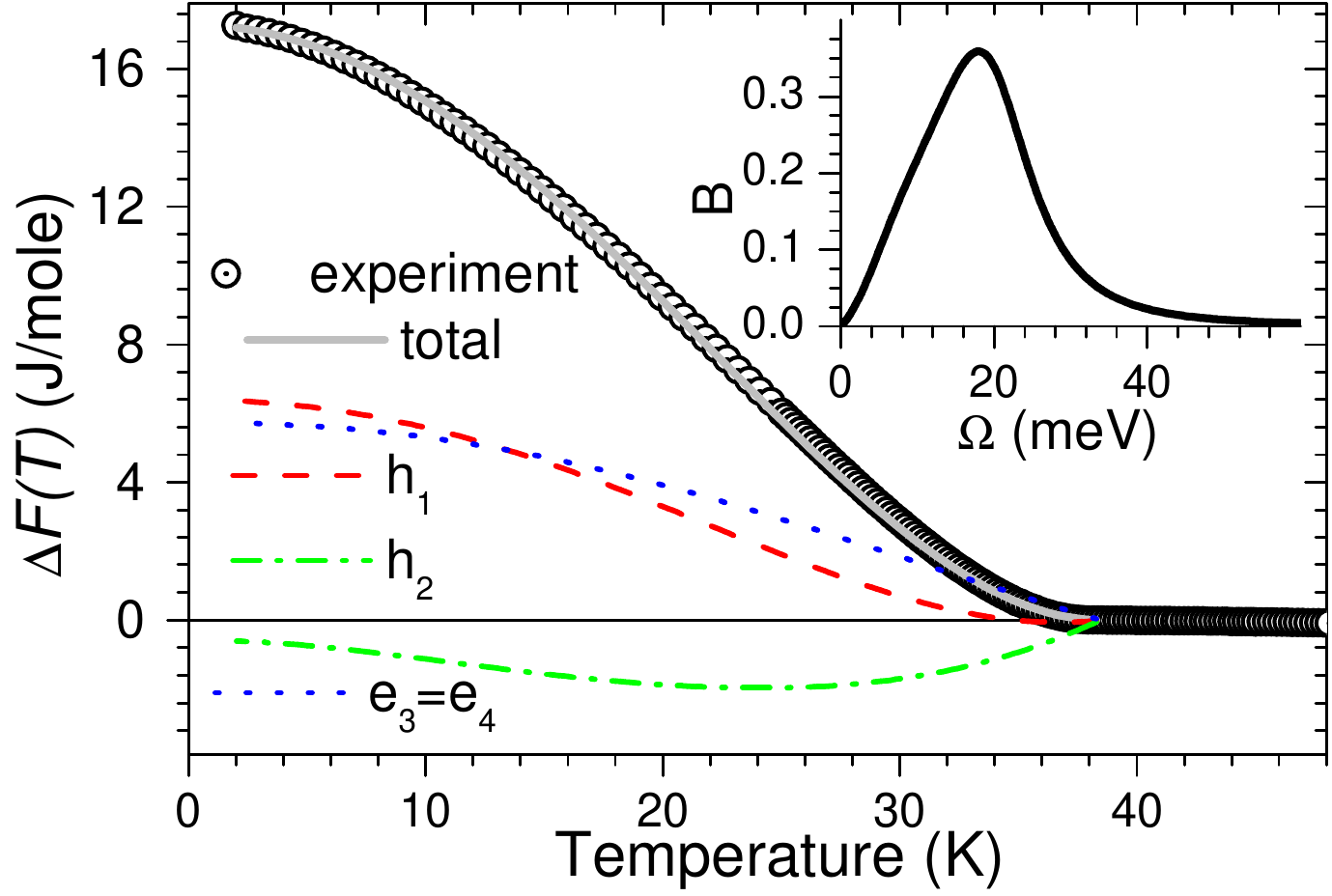}
\caption{Free-energy difference between normal and superconducting states, experimentally obtained (open circles) and calculated from
the four-band Eliashberg model (light gray line). The dashed, dash-dotted, and dotted lines represent the partial contributions of the individual bands. Inset: Spin-fluctuation coupling function $B(\Omega )$.}
\label{fig4}
\end{figure}

Finally, we illustrate that the strong coupling scenario can be described by fully microscopic calculations  in the framework of a spin-fluctuation model. Such models generically yield $s$-wave superconducting gaps with different signs on different bands (``$s_{\pm}$ state") \cite{Mazin}. It was shown in Refs. \onlinecite{Dolg2,Dolg3} that strong-coupling effects lead to merging of the gaps in a simple two-band $s_{\pm}$ model.
Therefore, a more realistic four-band model based on a band-structure with two hole bands and two electron bands crossing the Fermi surface needs to be considered \cite{Benfatto}.
The main input into the Eliashberg equations is the spectral function of the intermediate bosons. Following Ref. \onlinecite{Parker} we took a spin-fluctuation
coupling function $\tilde{B}_{ij}(\Omega )=\lambda _{ij}B(\Omega)$  [inset of Fig. 4] with a linear $\Omega$-dependence at low frequencies,   a   maximum at $\hbar \Omega^{\rm max}_{SF}$ = 18 meV, and a fast decay at $\Omega > \hbar \Omega^{\rm max}_{SF}$, in qualitative agreement with recent experimental data on the normal-state dynamical spin susceptibility of \BFCAX\ \cite{Inosov2,Lester}. Here $\lambda_{ij}$ is the coupling constant for pairing of electrons in bands $i$ and $j$. For our calculations we use the intraband coupling matrix elements $\lambda _{ii}$ = 0.2 in order to take account of the weak electron-phonon contribution \cite{Lilia},
 for interband repulsion $\lambda_{12}=\lambda_{34}$ = 0 due to the symmetry of the wave functions, and $\lambda _{13}=\lambda_{14}$ = $-$1.0, $\lambda
_{23}=\lambda _{24}$ = $-$0.2 ($\lambda _{ji}=\lambda _{ij}N^{h,e}_{i}(0)/N^{h,e}_{j}(0)$).
The corresponding densities of states are taken as $N^h_{1}(0)$ = 29 Ry$^{-1}$, ${N}^h_{2}(0)$ = 43 Ry$^{-1}$ for the hole bands, and as ${N}^e_{3}{(0) = N}^e_{4}(0)$ = 8.5 Ry$^{-1}$ for the two equivalent electron bands \cite{EPAPS}. The chosen parameters, $\lambda _{ij}(i\neq j)$ and $N^{h,e}_i(0)$, allow the best simultaneous fit to the experimental values of $T_c$, the superconducting gaps, and the temperature dependence of the free-energy difference, yielding $T_{c}$ = 38.5 K, and the following gap values: $\Delta^{h} _{1}$ = $-$8.5 meV, $\Delta^{h}_{2}$ = $-$3.6 meV for hole bands, and $\Delta^{e}_{3}=\Delta^{e} _{4}$ = 9.2 meV for electronic ones. The effective coupling constant averaged over all bands, $\lambda^{av}=\sum_{ij}N^{h,e}_i(0)\lambda_{ij}/N_{tot}$,
has a value of 1.9, remarkably close to the coupling constant reported for Pb$_{0.8}$Bi$_{0.2}$ \cite{Carbotte} and  in agreement with the conclusions of our phenomenological analysis presented in Fig. 3. The consistency with the experiment tolerates some variation of the parameters and remains satisfactory for $\hbar\Omega^{\rm max}_{SF}$ within the 10 - 20 meV energy range \cite{EPAPS}. A shift of $\hbar \Omega^{\rm max}_{SF}$ to lower energy leads to a larger values of $\lambda ^{av}$ with  $N^{h,e}_i(0)$  approaching the results of density functional theory (DFT) \cite{Sasha}. This accounts for the strong renormalization of the Sommerfeld constant $\gamma^N_K/\gamma_{DFT}\sim 5$.    

The superconductivity-induced
free-energy difference, $\Delta F(T)$, was calculated by using the expressions in Ref. \onlinecite{Dolgov}. 
The result is presented in Fig. 4, along with the partial band contributions to $\Delta F(T)$ which demonstrate that the superconductivity in the second hole band has an induced origin. Fig. 4 also shows that the result of the model calculation is in fairly good agreement with the free-energy difference obtained by integrating the experimentally measured $\Delta C_{el}(T)$, while some deviations between the calculated and observed specific heat can be seen in Fig. 2. Some such deviations are expected because feedback effects of superconductivity on the bosonic spectral function, which lead to the formation of a temperature-dependent ``resonant mode" in the spin fluctuation spectrum below $T_c$ \cite{Inosov2}, have not been considered in the calculations.

In summary, the specific heat of hole-doped \BKFA\ was measured on crystals of superior quality and phase purity. Our $\alpha$-model fit to the electronic part of the specific heat shows that there are two groups of superconducting gaps with magnitudes of 11 and 3.5 meV. The magnitude and shape of the superconducting jump anomaly indicate strong coupling with intermediate bosons, in good agreement with an Eliashberg analysis of a spin-fluctuation model.

\begin{acknowledgments}
We thank L. Boeri and A. N. Yaresko for helpful discussion and G. Siegle for experimental assistance. We acknowledge support by the Deutsche Forschungsgemeinschaft (DFG) via grant BO 3537/1-1 in SPP 1458.
\end{acknowledgments}


\newpage
\begin{widetext}

\begin{center}
\bf{SUPPLEMENTARY MATERIAL}
\end{center}

\begin{itemize}
\item\ 
\textbf{Evaluation of the normal-state specific heat at low temperature}
\end{itemize}
\begin{figure}[b!]
\begin{minipage}[b]{0.49\linewidth}
\centering
\includegraphics[width=7.8cm]{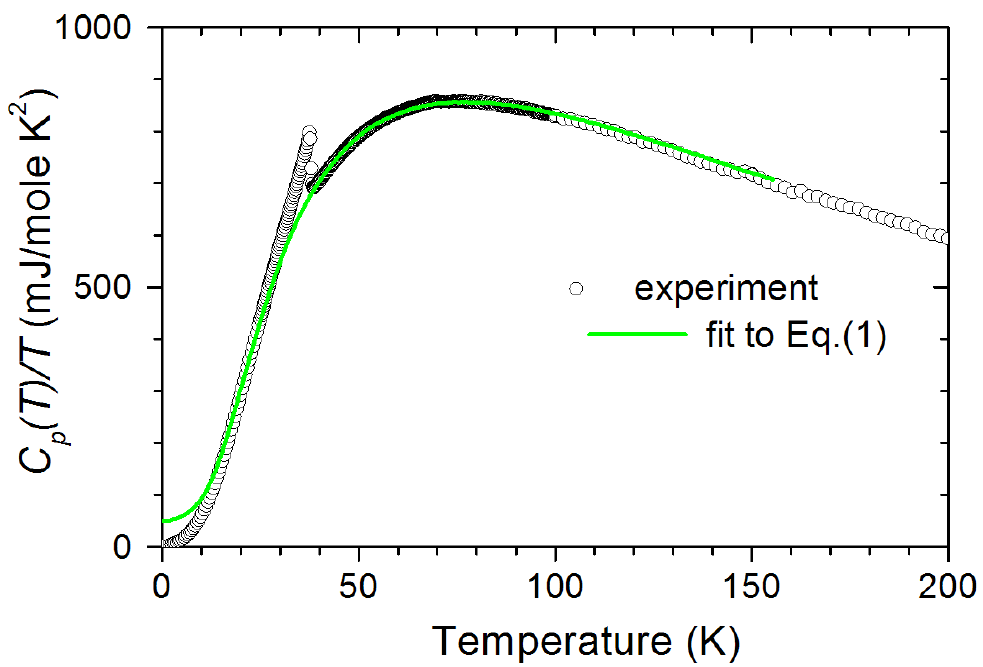}
\renewcommand{\thefigure}{S1}
\caption{Temperature dependence of (circles) the specific heat $C_p(T)/T$ of \BKFA\ and (green curve) its normal-state counterpart resulting from the corresponding state approximation by using the lattice specific heat of \BFMA \ as a reference, after Eq. (1) of the Letter.}
\end{minipage}
\hspace{0.2cm}
\begin{minipage}[b]{0.49\linewidth}
\centering
\includegraphics[width=7.8cm]{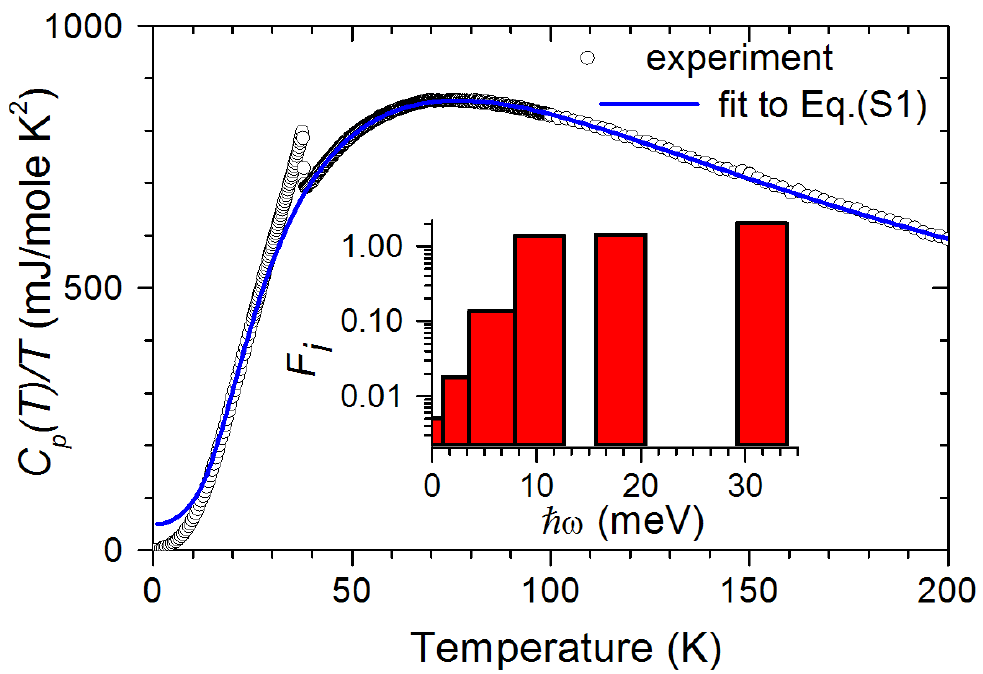}
\renewcommand{\thefigure}{S2}
\caption{Temperature dependence of (circles) the specific heat $C_p(T)/T$ of \BKFA\ and (blue curve) its normal-state counterpart resulting from    the simulation of  the lattice specific heat on a basis of Einstein modes, after Eq. (S1).  \textit{Inset:} The $\delta$-functions $F_i\delta(\omega - \omega_i)$ of the PDOS used for the fit are presented by a histogram.}
\protect\label{fig1}
\end{minipage}
\end{figure} 

\begin{table}[b!]
\centering
\renewcommand{\thetable}{SI}
\caption{Frequencies and weights of the six Einstein modes fit to the lattice specific heat of Ba$_{0.68}$K$_{0.32}$Fe$_2$As$_2$ ($\gamma$ = 49 \mJ).}
\label{tab-E}
\begin{tabular}{ccc}
\hline\hline
\ \ \ \\ \ \ \ \ \ $i$ \ \ \ \ \ \    & \ \ \ \ \ \ \ \ \ \ \ \ \ \ $\hbar \omega_i$ , meV \ \ \ (K) \ \ \ \ \ \ \ \ \ \ \ \ \ \ \ \ &  $F_i$ \ \ \ \ \ \\ \hline
1 & 1.92\ \ \ (22.3) &  0.005 \\ 
2 &  3.36\ \ \  (39.0) & 0.018 \\ 
3 & 5.89\ \ \  (68.4) &  0.136 \\ 
4 & 10.30\ \ \  (119.6) & 1.380 \\ 
5 & 18.04\ \ \  (209.4) & 1.450 \\ 
6 & 31.57\ \ \  (366.4) & 2.020\\ 

\hline\hline
\end{tabular}%

\end{table}

\begin{figure}[t!]
\centering
\renewcommand{\thefigure}{S3}
\includegraphics[width=8cm]{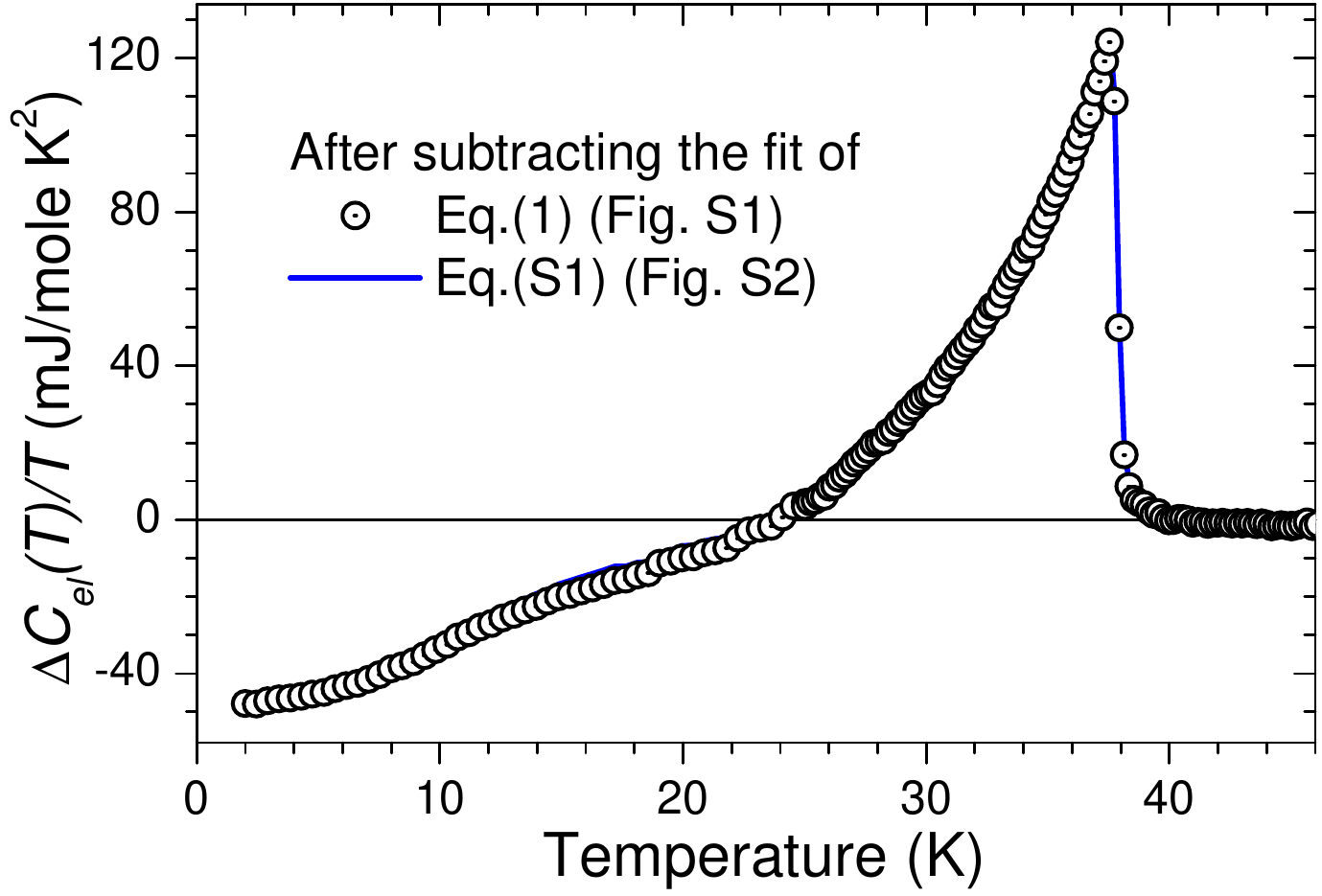}
\caption{Superconductvity-induced specific heat difference of of Ba$_{0.68}$K$_{0.32}$Fe$_2$As$_2$ obtained by subtractiongg the fit (green curve) by using the lattice specific heat of \BFMA \ as a reference, after Eq. (1) of the Letter and Fig. S1, and (blue curve) by simulating the lattice specific heat on a basis of Einstein modes, after Eq. (S1) and Fig. S2.}
\protect\label{figS3}
\end{figure}

In order to reliably extract the electronic contribution, $C_{el}(T)$, from the total measured specific heat, the contribution of lattice excitations, $C_{latt}(T)$, has to be accurately determined. The conventional method to evaluate the lattice specific heat of a superconductor is to recover the normal state by applying a magnetic field. This is not feasible for \BKFA, because the requisite field range is experimentally inaccessible. In our work, by using the lattice specific heat of \BFMA, $C^{Mn}_{latt}(T)$, as a reference, we evaluated $C^K_{latt}(T)$ of \BKFA \  as $A\cdot C_{latt}^{Mn}(B\cdot T)$, with scaling factors $A$ and $B$ of 0.95 and 1.03, respectively. The green curve in Fig. S1 represents the resulting normal-state specific heat of \BKFA \ from a least-squares fit of our data (open circles) to Eq.(S1) within the temperature range 40 - 160 K under the constraint of entropy conservation.

Alternatively, we also simulated $C^K_{latt}(T)$ by representing phonon density of states (PDOS) $F(\omega)$ by six $\delta$-functions $F_i\delta(\omega - \omega_i)$ on a basis of Einstein frequencies in a geometrical ratio $\omega_{i+1}=1.75\cdot \omega_i$. The corresponding normal-state specific heat is given by  
\begin{equation}
\renewcommand{\theequation}{S1}
C^K_{tot}(T) =C^K_{latt}(T)+\gamma_K^N\cdot T= 3 R \sum_{i=1}^6 F_i \frac{x^2_ie^{x_i}}{(e^{x_i}-1)^2}+\gamma_K^N\cdot T, \ \ \ \sum_{i=1}^6 F_i= 5   \label{Eins}
\end{equation}  
where $R$ is the universal gas constant,${\ x_i}=\hbar\omega_i/k_BT$, and the sum of the weights $F_i$ is constrained to the number of atoms per unit cell. A justification of this  trial PDOS is addressed, e.g., in Refs. 
S1-S3. The number of modes was chosen to be small enough to ensure the stability of the solution. The geometric ratio 1.75:1 allowed us to avoid spurious oscillations of the phonon DOS. The weights $F_i$ , the starting Einstein frequency $\omega_1$, and the Sommerfeld constant $\gamma_K^N$ ($\approx$ 49 \mJ) were determined by a least-squares fit of the high temperature (40 K $<T<$ 200 K) specific heat of \BKFA \  (open circles in Fig. S2) to Eq.(S1) under the constraint of entropy conservation. The results of the fit is given in Table SI. The $\delta$-functions of the phonon DOS $F(\omega)$ are represented by a histogram in the inset of Fig. S2. The frequencies $\hbar\omega_i$ are ranging from 2 to 32 meV with the main contributions at about 10, 18 and 32 meV, that is in a very good agreement with measured by inelastic x-ray scattering and calculated phonon frequencies in doped and undoped double-layered \BFA compounds [S4]. 
The blue curve in Fig. S2 represents the resulting normal-state counterpart of the specific heat of \BKFA  . Figure S3 compares results for the superconductivity-induced specific heat, $\Delta C_{el}(T)/T$,   obtained as the difference between the measured specific heat $C_{p}(T)/T$ of \BKFA \ and the normal-state specific heat resulting from two alternative approaches:  by using the lattice specific heat of \BFMA \ as a reference (open circles)  or by  simulating  the lattice specific heat on a basis of Einstein modes (blue curve). The difference between these two curves are within the uncertainty represented by the symbol size. 

\begin{itemize}
\item\ 
\textbf{Model spin-fluctuation spectral function, band and coupling parameters}
\end{itemize}

\begin{table}[b!]
\centering
\renewcommand{\thetable}{SII}
\caption{Microscopic paremeters of the four-band spin-fluctuation model with $\hbar \Omega^{\rm max}_{SF}$ = 18 meV and 13 meV fit the temperature dependence of the free-energy difference and yielding the corresponding $T_c$ and gap $\Delta^{h,e}_i$ values. $N^{h,e}_i$ is the density of states in the $i^{th}$ band, $\hat{\lambda}$ is the four-band matrix of the coupling constants, and MSE is the mean square error of the fit.  
}
\label{tab-P}
\begin{tabular}{cccc}
\hline\hline
\ \ \ \ \ \ \ \ \ \ \ \ \ \ \ \    & \ \ \ \ \  \ \ \ \ \ \ \ \ $\hbar \Omega^{\rm max}_{SF}$ = 18 $meV$ \ \ \ \ \ \ \ \ \ \ \ \ \ \ \ \ &  $\hbar \Omega^{\rm max}_{SF}$ = 13 $meV$\ \ \ \ & DFT calculations [S11]\ \ \\ \hline

$\hat{\lambda}$ &

$\left( 
\begin{array}{cccc}
0.2& 0 & -1.0  & -1.0  \\ 
0&0.2  & -0.2 & -0.2  \\ 
-3.41 & -1.01  & 0.2 &0  \\ 
-3.41 & -1.01  & 0 &0.2 
\end{array}%
\right)$
&
$\left( 
\begin{array}{cccc}
0.2& 0 & -1.7  & -1.7  \\ 
0&0.2  & -0.25 & -0.25  \\ 
-5.34 & -0.89  & 0.2 &0  \\ 
-5.34 & -0.89  & 0 &0.2 
\end{array}%
\right)$
&
\
\\ 
($N^h_1,N^h_2,N^e_1,N^e_2$), Ry$^{-1}$ & (29\ \ 43\ \ 8.5\ \ 8.5) & (22\ \ 25\ \ 7.0\ \ 7.0) & (26\ \ 20\ \ 6.6\ \ 6.6)\\ 
$\lambda^{av}$ & 1.89\ \ \  &  3.06 \\ 
($\Delta^h_1,\Delta^h_2,\Delta^e_1,\Delta^e_2$), meV & (-8.5\ \ -3.6 \ \ 9.2 \ \ 9.2)& (-9.5\ \ -3.6 \ \ 9.8 \ \ 9.8)\ \ \\ 
$T_c$, K & 
38.5 & 39.2 \\ 
MSE, (J/mol)$^2$ & 0.015 & 0.034\\
\hline\hline  
\end{tabular}%
\end{table}

In our Letter, we describe the results of a fully microscopic calculation of the
specific heat in the framework of a multiband spin-fluctuation model. The main input into the
Eliashberg equations is the spectral function of the intermediate bosons. Early calculations of magnetic pairing in the high-$T_c$ cuprate superconductors employed two alternative parametrized models of the susceptibility: The analysis
of Radtke \etal \ [S5] 
[Radtke-Ullah-Levin-Norman (RULN) model] invoked neutron-scattering measurements, while that of Millis \etal \ [S6] 
[Millis-Monien-Pines (MMP) model] was based on NMR data. Because of a long tail at high frequencies  $\propto1/\omega$, the MMP spectrum leads to the divergence of the first momentum (proportional to the Hopfield-factor $\eta$) and the overestimation of $T_{c}$
and superconducting gaps. Very recently, applying the   MMP spectrum for spin fluctuations, the authors of Ref. S8 
came to the conclusion about weak (intermediate) coupling in pnictides. The necessity of an improvement of the high energy
dependence of the MMP spin-fluctuation spectral function was discussed in
Ref. S7. 
\begin{figure}[t!]
\renewcommand{\thefigure}{S4}
\includegraphics[width=7.8cm]{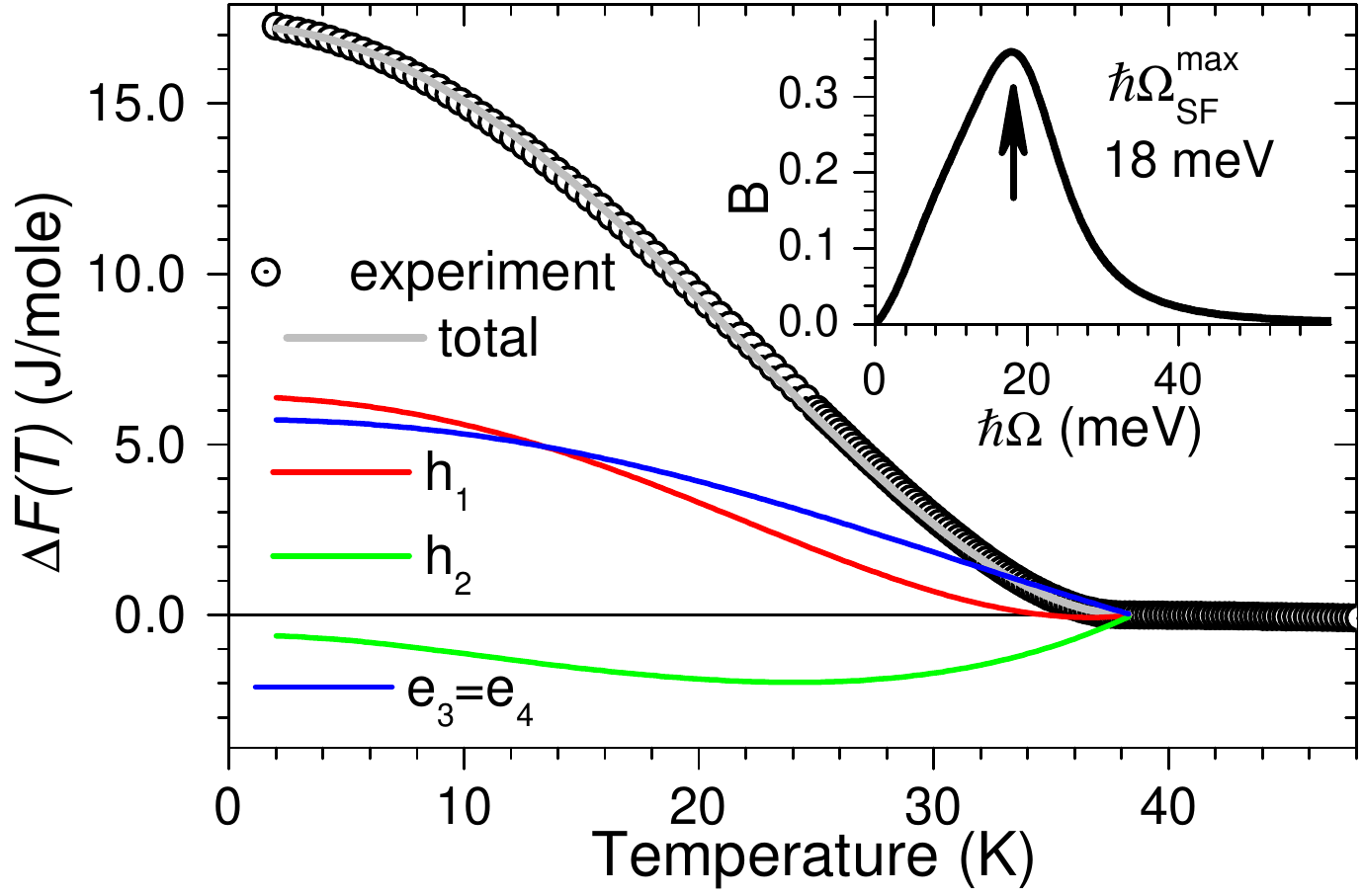}
\includegraphics[width=7.8cm]{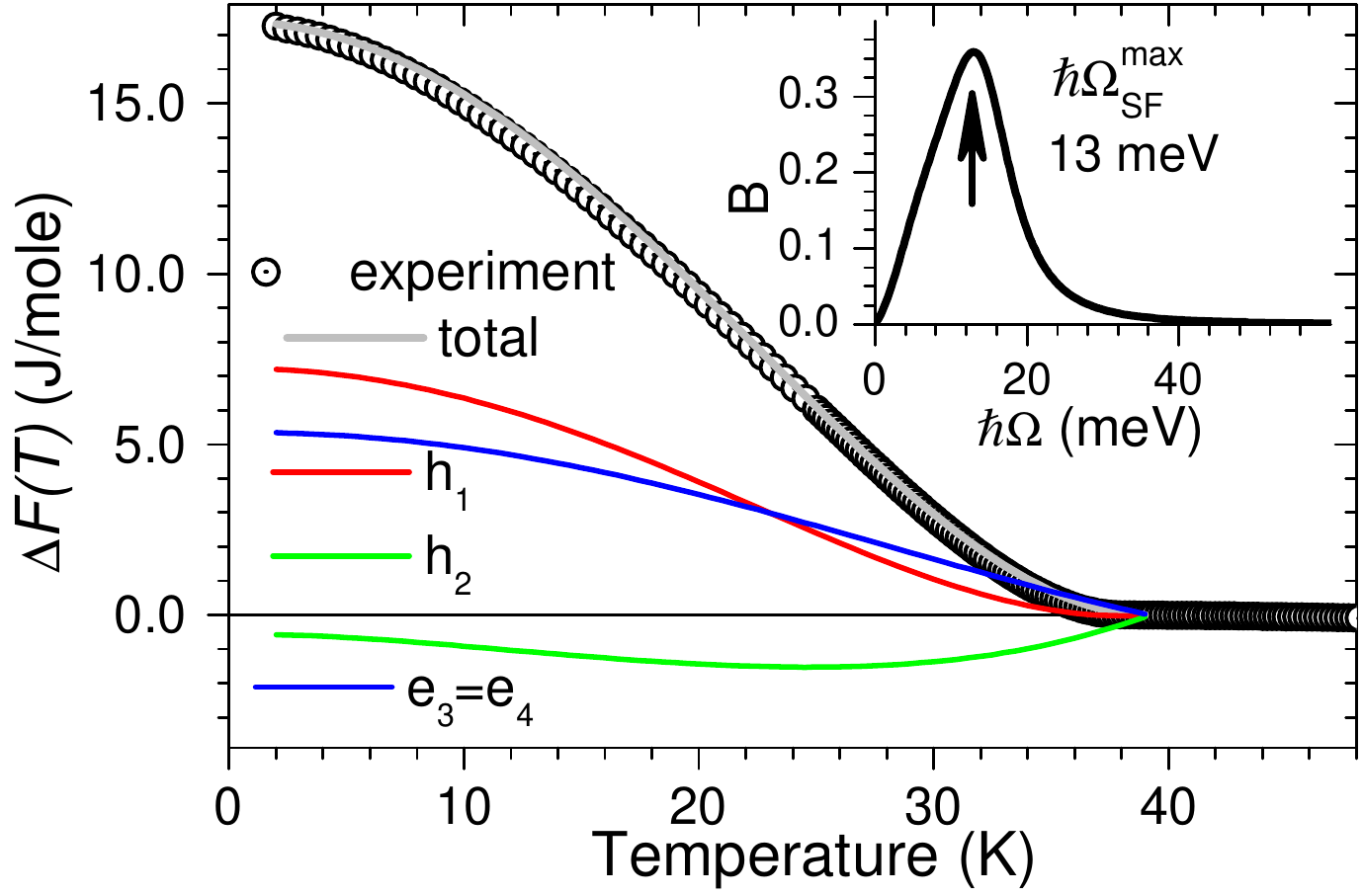}
\caption{The best fit to the free-energy difference of the four-band Eliashberg model with the coupling function $B(\Omega)$ peaked at $\hbar \Omega^{\rm max}_{SF}$ = 18 meV (\textit{left}) and 13 meV (\textit{right}) and parameters listed in Table SII.. The red, green, and blue  lines represent the partial contributions of the individual
bands.}
\label{Gap18}
\end{figure} 
\begin{figure}[t!]
\renewcommand{\thefigure}{S5}
\includegraphics[width=7.8cm]{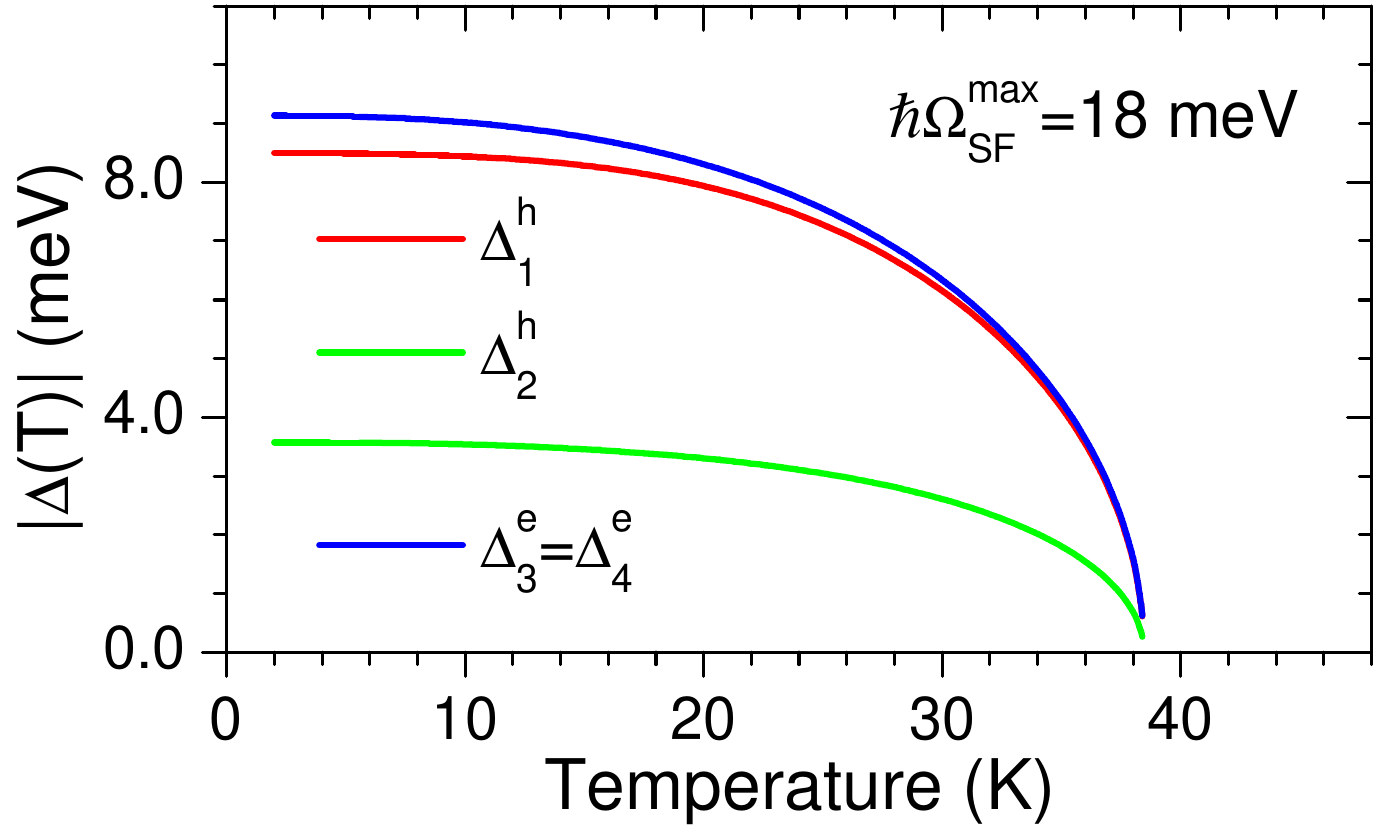}
\includegraphics[width=7.8cm]{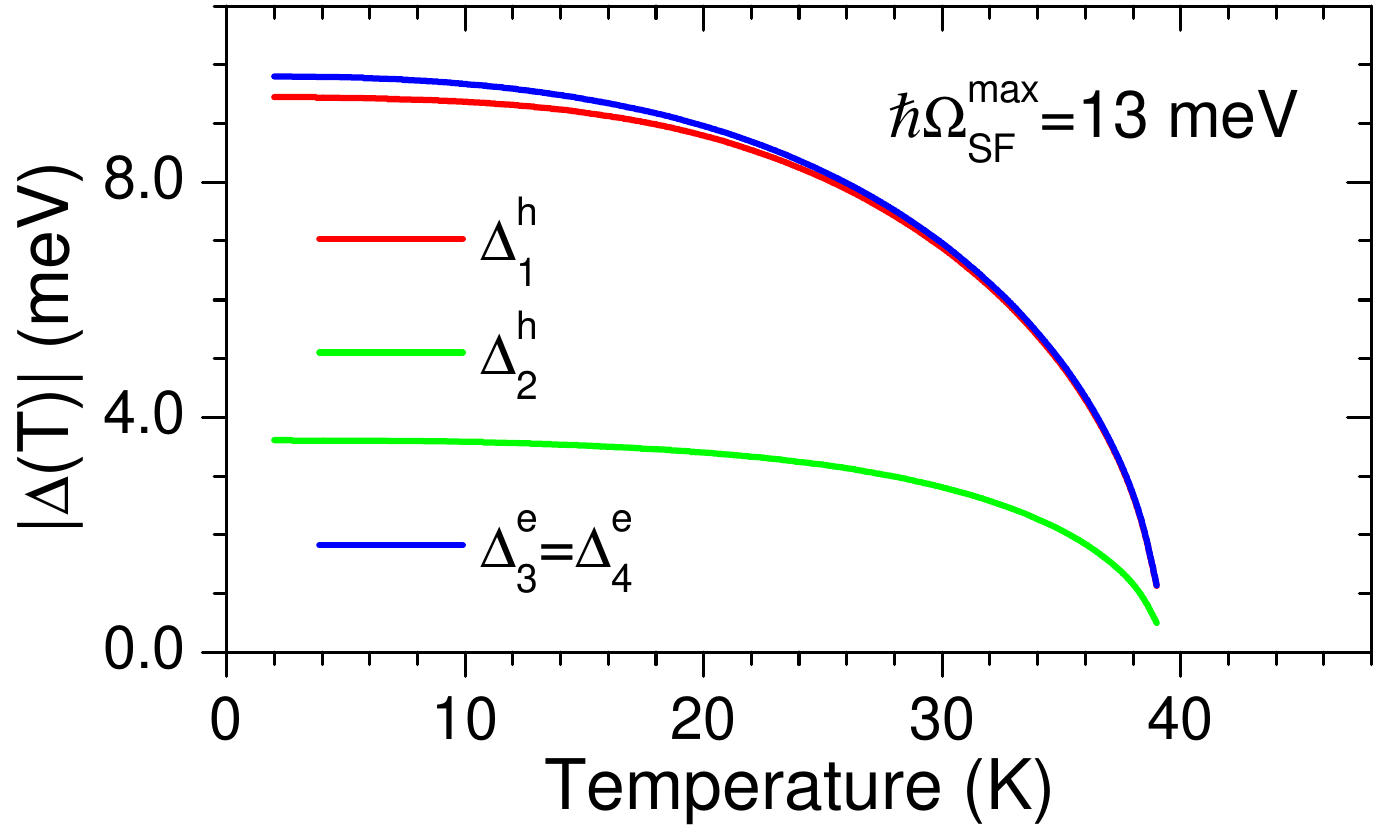}
\caption{Temperature dependence of the superconducting energy gaps calculated from the four-band Eliashberg model with the coupling function $B(\Omega)$ peaked at $\hbar \Omega^{\rm max}_{SF}$ = 18 meV (\textit{left}) and 13 meV (\textit{right}) and parameters listed in Table SII.}
\label{Gap13}
\end{figure} 
In our analysis we took a more realistic coupling
function based on neutron-scattering measurements, with a linear $\Omega$-dependence at low frequencies,   a   maximum at $\hbar \Omega^{\rm max}_{SF}$ = 18 meV, and a fast decay at $\Omega > \hbar \Omega^{\rm max}_{SF}$, in qualitative agreement with recent experimental data on the normal-state dynamical spin susceptibility of \BFCAX\ [S9, S10].
With this spectral function shown in the insert of Fig.4 (also in Fig.S4) and the variable parameters described and listed in the Letter (also in  Table SII), we obtained the best simultaneous fit to the experimental values of $T_c$, the superconducting gaps, and the temperature dependence of the free-energy difference.
The effective coupling constant averaged over all bands, $\lambda^{av}=\sum_{ij}N^{h,e}_i(0)\lambda_{ij}/N^{h,e}_{tot}\approx 1.9$, is remarkably close to the coupling constant of Pb$_{0.8}$Bi$_{0.2}$ and  confirms our conclusion about strong coupling in \BKFA\ (see Fig.3 of the Letter for  comparison). The consistency with the experiment tolerates some variation of the parameters and remains satisfactory for $\hbar\Omega^{\rm max}_{SF}$ within the 10 - 20 meV energy range. 
In Table SII and Figs. S4-S5 we compare our results for two representative coupling functions with $\hbar \Omega^{\rm max}_{SF}$ = 18 meV and 13 meV, illustrating that a shift of $\hbar \Omega^{\rm max}_{SF}$ to lower energy leads to a larger values of $\lambda ^{av}$  with corresponding
densities of states $N^{h,e}_i(0)$ approaching the first-principle Density Functional results that accounts for the strong renormalization of the Sommerfeld constant $\gamma^N_K/\gamma_{DFT}\sim 5$.    

\begin{table}[ht]
\begin{tabular}{ll}
\verb=[S1]= J.W. Loram, J. Phys. C \textbf{19}, 6113 (1986).   & \verb=[S7]= H.-B. Sch\"uttler, M.R. Norman, Phys. Rev. B\textbf{ 54}, 13295 (1996). \\
\verb=[S2]= J.W. Loram \textit{et al}., Physica C \textbf{168}, 47 (1990).&\verb=[S8]= L. Benfatto et al., Phys. Rev. B\textbf{ 80}, 214522 (2009).  \\
\verb=[S3]= Y. Wang \textit{et al.}, Physica C \textbf{355}, 179 (2001). & \verb=[S9]= D. S. Inosov et al., Nature Phys.  \textbf{6}, 178 (2010)\\ \verb=[S4]= D. Reznik \textit{et al.}, Phys. Rev. B \textbf{80}, 214534 (2009).&  \verb=[S10]= C. Lester et al., Phys. Rev. B \textbf{81}, 064505 (2010).\\ \verb=[S5]= R. J. Radtke \textit{et al.}, Phys. Rev. B \textbf{46}, 11975 (1992). & \verb=[S11]= A. N. Yaresko, private communication.\\ 
\verb=[S6]= A. Millis, H. Monien, D. Pines, Phys. Rev. B \textbf{42}, 167 (1990).  \\
\end{tabular}
\end{table}
\end{widetext}

\end{document}